# Role of the Nephelauxetic Effect in Engineering $Mn^{4+}$ Luminescence Kinetics for Lifetime-Based Thermometry

A. Basheer[1], M. Szymczak[1], M. Piasecki[2], A. M. Srivastava[3], M.G. Brik[2,4,5,6,7], L. Marciniak[1*]

[1] *Institute of Low Temperature and Structure Research, Polish Academy of Sciences, Okólna 2, 50-422 Wrocław, Poland*

[2] *Faculty of Science and Technology, Jan Długosz University, Armii Krajowej 13/15, 42-200 Częstochowa, Poland*

[3] *Current Chemicals, Inc., 1099 Ivanhoe Road, Cleveland, Ohio 44110, United States of America*

[4] *School of Integrated Circuits, Chongqing University of Posts and Telecommunications, Chongqing 400065, China;*

[5] *Center of Excellence for Photoconversion, Vinča Institute of Nuclear Sciences, University of Belgrade, 11001 Belgrade, Serbia;*

[6] *Institute of Physics, University of Tartu, 50411 Tartu, Estonia;*

[7] *Academy of Romanian Scientists, 050044 Bucharest, Romania;*

\* corresponding author: l.marciniak@intibs.pl;

**Abstract**

Although the considerable potential of luminescence thermometry based on emission kinetics has been widely demonstrated, reliable tools for the intentional prediction of thermometric performance remain limited. To address this challenge, the present work introduces an approach that enables a theoretical description of the $^2$E-state lifetime of $Mn^{4+}$ ions, as well as the absolute and relative sensitivities, in terms of the nephelauxetic effect within the group of double perovskites: $Sr_2InNbO_6$, $Sr_2InTaO_6$, $Ba_2InTaO_6$, and $Ba_2InNbO_6$. Our results clearly show that, contrary to common assumptions, the *Dq/B* ratio is not the primary factor governing either the spectroscopic behavior of $Mn^{4+}$ ions or the thermometric performance of $Mn^{4+}$-doped phosphors. Instead, the nephelauxetic *β$_1$* parameter plays the dominant role. The empirical analysis carried out in this study led to the development of a predictive model that enables estimation of *S$_{AMAX}$* and *S$_{RMAX}$* values based exclusively on *β$_1$*. This methodology represents a significant step toward the rational design of lifetime-based luminescence thermometers with predefined thermometric characteristics tailored to the requirements of specific applications.

**Introduction**

The strong temperature dependence of luminescent properties in phosphor materials underpins the principle of luminescent thermometry, enabling remote, two-dimensional, and even three-dimensional thermal imaging[1–7]. Among the various spectroscopic parameters available for temperature readout, the luminescence intensity ratio (*LIR*) and luminescence kinetics are the most widely employed because of their reliability and practical simplicity[8–14]. However, the shape of the emission spectrum and consequently the calculated *LIR* can be significantly distorted by the optical characteristics of the medium between the thermometer and the detector, particularly in absorbing or scattering environments[13,15]. Such distortions compromise the accuracy of temperature determination. In contrast, luminescence kinetics is inherently immune to these spectral distortions, making it an especially robust alternative. For lifetime-based thermometry, achieving high relative sensitivity typically requires a steep, thermally activated decrease in the luminescence lifetime. However, such steep changes often limit the thermal operating range of the thermometer. Conversely, applications that demand a broad thermal operating range benefit from a gradual, monotonic lifetime variation. Therefore, designing effective luminescent thermometers requires the ability to tailor their thermometric performance to meet the specific requirements of different applications.

In phosphors doped with lanthanide ions, the dominant process responsible for thermal depopulation of the emitting states is multiphonon relaxation[16–19]. This process is governed by the phonon energy of the host lattice and the energy gap between the emitting level and the lower-lying states[20]. While the phonon energy of the host can be partially tuned via compositional modification, the energy levels of lanthanide ions, due to their shielded 4$f$ orbitals, are only weakly perturbed by the crystal field environment[17,21,22]. Consequently, the energy separations between levels remain essentially constant across different host materials. An alternative strategy involves co-doping with additional ions to introduce energy transfer pathways that enhance thermal sensitivity[11,12,23,24]. However, such approaches may introduce unwanted side effects, such as concentration quenching or spectral overlap[25,26]. As a result, phosphors doped with lanthanide ions offer limited flexibility in tuning the thermometric performance of lifetime-based luminescence thermometers.

In contrast, phosphors doped with transition metal ions exhibit high sensitivity to both temperature changes and local crystal field variations[27–30]. In such systems, even small changes in the crystal field strength and bonding covalence can induce substantial modifications in spectroscopic behavior [31–37]. A particularly illustrative example is the $Mn^{4+}$ ion ($3d^3$ electronic configuration), for which the crystal field strength parameter $Dq/B$ depends on the metal-oxygen bond length ($R$) according to the approximate relationship $Dq/B \sim R^{-5}$ [28,38,39]. Therefore, adjusting the chemical composition of the host material, and thereby modifying $R$, directly affects the thermal activation energy that governs the temperature-dependent luminescence decay from the $^2$E state[27,40]. This provides a route to broadly tune the thermometric behavior of Mn$^{4+}$-based thermometers. Nevertheless, *Dq/B* is not the only factor influencing the thermal evolution of Mn$^{4+}$ luminescence kinetics. For this reason, the rational design of application-specific luminescent thermometers requires systematic investigation of the correlations between structural parameters and chemical composition of the host lattice and the spectroscopic behavior of Mn$^{4+}$-doped phosphors.

To address this issue, the present work systematically examines the luminescence of Mn$^{4+}$ ions in four different double perovskite hosts, Sr$_2$InNbO$_6$, Sr$_2$InTaO$_6$, Ba$_2$InNbO$_6$, and Ba$_2$InTaO$_6$. By systematically varying both the *A*-site and *B'*-site cations, we aim to establish structure-luminescent property correlations that govern the relative sensitivity and overall thermometric performance of these Mn$^{4+}$-doped phosphors.

**Experimental**

*Synthesis*

Polycrystalline powders of Sr$_2$InNbO$_6$:1%Mn$^{4+}$, Sr$_2$InTaO$_6$:1%Mn$^{4+}$, Ba$_2$InNbO$_6$:1%Mn$^{4+}$, and Ba$_2$InTaO$_6$:1%Mn$^{4+}$ were synthesized via high temperature solid-state reaction route. Strontium(II) carbonate SrCO$_3$, indium(III) oxide In$_2$O$_3$, tantalum(V) oxide Ta$_2$O$_5$, niobium(V) Oxide Nb$_2$O$_5$, barium(II) carbonate BaCO$_3$, manganese(II) chloride tetrahydrate MnCl$_2$·4H$_2$O were weighed stoichiometrically and thoroughly ground in an agate mortar in hexane medium. Then, the resulting mixtures were transferred to crucibles separately and were initially calcined for 6 h at 600 °C in a muffle furnace. After calcination, the precursor was reground and then subject to further calcination at 1300 °C for 6h.

*Characterization*

The obtained materials were examined by X-ray powder diffraction (XRD) using a PANalytical X'Pert Pro diffractometer in Bragg-Brentano geometry, using Ni-filtered Cu Kα radiation (V = 40 kV, I = 30 mA). Measurements were made in the range of 10-90°, and the acquisition time was approximately 30 min. The samples were also investigated using a scanning electron microscopy (SEM, FEI Nova NanoSEM 230) equipped with an energy-dispersive X-ray spectrometer (EDX, EDAX Apollo X Silicon Drift Correction) compatible with genesis EDAX microanalysis Software. The samples were dispersed in alcohol, and a drop of the suspension was placed onto a carbon stub. SEM images were then collected using an accelerating voltage of 5.0 kV, while EDS measurements were performed at a higher accelerating voltage of 30 kV.

Raman spectra were collected using an Edinburgh Instruments RMS-1000 Raman microscope equipped with a 532 nm excitation laser, a 20x microscope objective, and an 1800 lines mm$^{-1}$ diffraction grating. Temperature stabilization during the measurements was ensured using a temperature-control THMS600 stage from Linkam.

Excitation and emission spectra were measured using an FLS1000 spectrometer from Edinburgh Instruments, equipped with a 450 W xenon lamp as the excitation source, and an R928P side window photomultiplier tube from Hamamatsu as a detector. The luminescence decay curves were recorded on this spectrometer with a pulsed excitation, provided by a µFlash lamp. Temperature of the samples during the measurements was controlled using THMS600 heating-cooling stage from Linkam.

The luminescence decay profiles were fitted with a double exponential function, as expressed below:

$$I(t) = I_0 + A_1 \cdot \exp\left(-\frac{t}{\tau_1}\right) + A_2 \cdot \exp\left(-\frac{t}{\tau_2}\right) \qquad (1)$$

where $I(t)$, $A_1$, $A_2$, $I_0$, $\tau_1$, and $\tau_2$, represent the luminescence intensity at time $t$ right after laser pulse excitation, preexponential factors, offset, and decay parameters, respectively. Based on these parameters the average lifetime $\tau_{avr}$ were calculated as follows:

$$\tau_{avr} = \frac{A_1\tau_1^2 + A_2\tau_2^2}{A_1\tau_1 + A_2\tau_2} \quad (2)$$

**Results and discussion**

The structure of the $A_2BB'O_6$ double perovskites is derived from the simple perovskite through the ordered arrangement of the $B$ and $B'$ cations in alternating octahedral sites[31,41–45]. In this structure, the $A$ site is usually occupied by the larger cations - $Sr^{2+}$ or $Ba^{2+}$. Each $Sr^{2+}$ ion is surrounded by eight oxygen atoms, whereas each $Ba^{2+}$ ion is coordinated by twelve oxygen atoms (Figure 1a, b)[46–51]. On the other hand, the $B$ site is occupied by $In^{3+}$ ions, and $B'$ by $Ta^{5+}$ or $Nb^{5+}$, respectively, each exhibiting octahedral coordination[52–57]. A comparison of the structural parameters of the host materials examined in this study presented in Table 1 indicates that $Sr_2InNbO_6:Mn^{4+}$ and $Sr_2InTaO_6:Mn^{4+}$ crystallize in a monoclinic structure with a $P2_1/n$ space group, whereas $Ba_2InTaO_6:Mn^{4+}$ and $Ba_2InNbO_6:Mn^{4+}$ compositions adopt the cubic structure with $Fm\bar{3}m$ space group[52–55]. The symmetry of these host structures was confirmed by calculating the Goldschmidt geometric tolerance factor $t$, which for double perovskites is defined as follows[58]:

$$t = \left[\frac{r_A + r_O}{\sqrt{2}\left(\frac{r_B + r_{B'}}{2} + r_O\right)}\right] \quad (3)$$

where $r_A$ is the ionic radii of the $A$ cation, $r_B$ and $r_{B'}$ are the ionic radii of $B$ and $B'$ ions in six-fold coordination and $r_O$ is the ionic radii of the oxygen ion (1.40 Å). The value of $t$ reflects the stability of the perovskite structure. The highest stability occurs for $t = 1$, which corresponds with the ideal cubic perovskite structure. According to literature, when $t$ exceeds 0.98, the material typically adopts a cubic $Fm\bar{3}m$ ($a_0$, $a_0$, $a_0$) structure, whereas for $t$ values between 0.96 and 0.98, rhombohedral or tetragonal symmetries are most commonly reported[47,49,52,53,59–61]. The tolerance factor for the $Ba_2In(Ta,Nb)O_6$ compositions is close to 1 (Table 1), consistent with their adoption of a cubic structure. It implies linear $In^{3+}$-$O^{2-}$-$(Ta,Nb)^{5+}$ bonds ($180^0$) and the ideal $(NbO_6)^{7-}$ and $(TaO_6)^{7-}$ octahedra. The larger $Ba^{2+}$ ions are in a twelve-fold coordination in this double perovskite. In contrast, replacing $Ba^{2+}$ with the smaller $Sr^{2+}$ ion introduces a size mismatch between the *A*-site cation and the cuboctahedral cavity formed by the twelve surrounding oxide ions. This mismatch drives octahedral tilting, lowering the symmetry. With a calculated tolerance factor of $t$ =0.88, the $Sr_2In(Ta,Nb)O_6$ analogs are predicted to adopt a monoclinic structure, in agreement with X-ray diffraction results (Figure 1b). In these $Sr^{2+}$-based compounds, $Sr^{2+}$ is eight-fold coordinated, and the $In^{3+}$-$O^{2-}$-$(Ta,Nb)^{5+}$ bond angle decreases to approximately $169^0$, reflecting the octahedral tilting[46–49,52,57].

Interpreting the optical data requires identifying the specific host-lattice site occupied by $Mn^{4+}$. In these double perovskite structures, two distinct octahedral sites are available for substitution by $Mn^{4+}$ ions. Although one might argue that the coupled substitution $2Mn^{4+} \rightarrow In^{3+} + Nb^{5+}/Ta^{5+}$ is favourable from a charge-balance perspective, the ionic-radius mismatch strongly disfavours this mechanism, as discussed below. The suitability of this substitution can be assessed by calculating the radius percentage difference $D_R$ between $Mn^{4+}$ and host-cation ($Nb^{5+}/Ta^{5+}$, $In^{3+}$) using the following formula[62]:

$$D_R = \frac{R_m(CN) - R_d(CN)}{R_m(CN)} \times 100\% \qquad (4)$$

where $R_m(CN)$ and $R_d(CN)$ are the ionic radii of the host and dopant cation, respectively. The ionic radii of the ions in sixfold coordination considered in this case are as follows: $Mn^{4+}$ (0.53 Å), $In^{3+}$ (0.80 Å), $Nb^{5+}/Ta^{5+}$ (0.64 Å)[63]. Because favourable substitution typically requires the $D_R$ value to be below 30%, the calculated $D_R$ values of 33.75% for $Mn^{4+}$ substituting $In^{3+}$ and 17.2% substituting $Nb^{5+}/Ta^{5+}$ indicate a clear preference for incorporation at the $Nb^{5+}/Ta^{5+}$ sites. Substitutions at these sites minimize local elastic strain effects or lattice distortions. However, the substitution of $Nb^{5+}/Ta^{5+}$ by $Mn^{4+}$ ions, which differ in charge, is expected to disrupt local charge neutrality generating positively charged defects such as oxygen vacancies or anti-site disorder to restore balance[48,52]. This disruption not only affects the charge distribution but can also influence the crystal structure[64–66]. Furthermore, this structural distortion is consistent with reports of around 10% anti-site disorder, where $In^{3+}$ and $Nb^{5+}/Ta^{5+}$ exchange positions within the rock salt type structure, indicative of translational stacking faults[57,67].

**Table 1**. Comparison of structural parameters of $Sr_2InNbO_6$, $Sr_2InTaO_6$, $Ba_2InTaO_6$ and $Ba_2InNbO_6$ host materials.

|  | $Sr_2InNbO_6$ | $Sr_2InTaO_6$ | $Ba_2InTaO_6$ | $Ba_2InNbO_6$ |
|---|---|---|---|---|
| Crystal structure | monoclinic | monoclinic | cubic | cubic |
| Space group | $P2_1/n$ (14) | $P2_1/n$ (14) | $Fm\bar{3}m$ (225) | $Fm\bar{3}m$ (225) |
| Cell parameter (Å) | $a$ = 5.7351(2) | $a$ = 5.73356(10) | $a$ = 8.29640(1) | $a$ = 8.2819(2) |
|  | $b$ = 5.7416(2) | $b$ = 5.74052(10) |  |  |
|  | $c$ = 8.1102(3) | $c$ = 8.10905(14) |  |  |
| Cell volume (Å$^3$) | 267.058 | 266.898 | 571.043 | 568.054 |
| Average bond length ($Ta^{5+}/Nb^{5+}$-$O^{2-}$) (Å) | 2.003 | 2.0106 | 2.016 | 2.0705 |

In the fully ordered structure of the $A_2BB'O_6$ double perovskite oxides, each $B$ cation has only $B'$ as the next nearest neighbour and vice versa. However, this ordering is never complete and some degree of mixing between the $B$ and $B'$ cations occur. The degree of order is determined by the difference between the ionic size and charge of the $B$-site cations[68–70]. The degree of mixing increases with smaller ionic radii and charge differences between the two $B$-site cations.

According to Woodward et al. when the ionic-radius difference is $\Delta r \leq 0.260$ Å, complete (100%) *B/B'* ordering is generally not observed[71–73]. Further, when the *A*- site cation is smaller and the crystal symmetry is reduced, the same *B/B'* combination tends to show higher ordering levels. For example, $Sr_2ScNbO_6$ and $Ca_2ScNbO_6$, both adopting the monoclinic structure, exhibit ordering levels of 69% and 96%[74].

Applying these insights to the present compositions, the ionic radius difference between $In^{3+}$ and $Nb^{5+}/Ta^{5+}$ is $\Delta r = 0.16$ Å, a value that suggests a substantial degree of *B/B* disorder. This mixing is expected to be more pronounced in the cubic $Ba_2In(Ta,Nb)O_6$ phases than in the monoclinic $Sr_2In(Ta,Nb)O_6$ analogs, where lower symmetry typically favours higher ordering. A long-range order of 80% and 86% has been reported for $Sr_2InNbO_6$ and $Sr_2InTaO_6$, respectively[48,52]. Such considerations are essential when interpreting the spectroscopic features and luminescence lifetimes of $Mn^{4+}$ in these double-perovskite hosts and would be important when discussing the spectroscopic and the lifetime data of the $Mn^{4+}$ in these double perovskites in the later part of this work.

X-ray diffraction patterns of $Sr_2InTaO_6:Mn^{4+}$, $Ba_2InTaO_6:Mn^{4+}$, $Sr_2InNbO_6:Mn^{4+}$ and $Ba_2InNbO_6: Mn^{4+}$, presented in Figure 1c, correlate well with the standard reference data. No traces of the residual or impurities were detected, indicating successful incorporation of $Mn^{4+}$ into the host materials. The replacement of $Nb^{5+}/Ta^{5+}$ (*R*=0.64 Å) by smaller $Mn^{4+}$ ions (*R*=0.53 Å) is expected to cause lattice contraction, which should result in a shift of the reflections toward higher angles. However, this shift is not clearly observed in the investigated case, most likely due to the low concentration of $Mn^{4+}$ ions.

Room-temperature Raman spectra were collected for all compounds to probe their structural symmetry and vibrational properties (Figure 1d). The recorded spectra clearly reveal the presence of two distinct structural groups. The $Ba^{2+}$-based compounds exhibit Raman spectra with 14 active modes, consistent with a cubic crystal structure[57]. In contrast, the $Sr^{2+}$-based compounds exhibit a much larger number of Raman-active modes, indicative of the lower-symmetry monoclinic structure. For the $Sr^{2+}$-based compounds, the most intense Raman bands appear near 800, 560, 420 $cm^{-1}$, along with additional modes below 300 $cm^{-1}$. In contrast, the $Ba^{2+}$-based analogs are characterized by dominant low-frequency band around 100 $cm^{-1}$, and pronounced features near 400 $cm^{-1}$. The mode near 800 $cm^{-1}$ is assigned to the symmetric stretching vibrations of the $(TaO_6)^{7-}/(NbO_6)^{7-}$ octahedra. Bands in the 500-600 $cm^{-1}$ region arise from $Nb^{5+}/Ta^{5+}$-$O^{2-}$ stretching modes, including antisymmetric stretching vibrations. Vibrations observed between 300 and 450 $cm^{-1}$ correspond to $O^{2-}$-$Nb^{5+}/Ta^{5+}$-$O^{2-}$ bending modes. At lower wavenumbers (<300 $cm^{-1}$), the spectra are dominated by lattice translational and vibrational modes [57,67,75,76]. A comparative analysis of the Raman spectra reveals systematic differences arising from substitutions at both the *A* and *B* cation sites in these double perovskites. For compounds containing the same *B'* cation, the $Ba^{2+}$-based members consistently exhibit Raman bands shifted to lower wavenumbers most notably the mode near 400 $cm^{-1}$. This shift reflects the larger ionic radius of $Ba^{2+}$ relative to $Sr^{2+}$, which weakens *A*-O interactions, reduces bond stiffness, and lowers the corresponding vibrational frequencies. In contrast, the smaller $Sr^{2+}$ ion strengthens the ionic character of the $Sr^{2+}$-$O^{2-}$ bonds and contracts the lattice, producing a systematic shift of Raman modes toward higher wavenumbers.

Substitution at the *B'* site ($Ta^{5+} \rightarrow Nb^{5+}$) produces more subtle changes in the Raman spectra. Because $Ta^{5+}$ and $Nb^{5+}$ possess identical ionic radii and the same charge, the overall crystal symmetry remains essentially unchanged. The subtle variations observed in the spectra therefore arise mainly from differences in $Ta^{5+}$-$O^{2-}$ versus $Nb^{5+}$-$O^{2-}$ bond strengths, which produce slight shifts in the positions and relative intensities of the octahedral stretching and bending modes. The analysis of the SEM images for the analyzed phosphors revealed that they consist of microcrystalline and aggregated crystals (Figure 1 e-h).

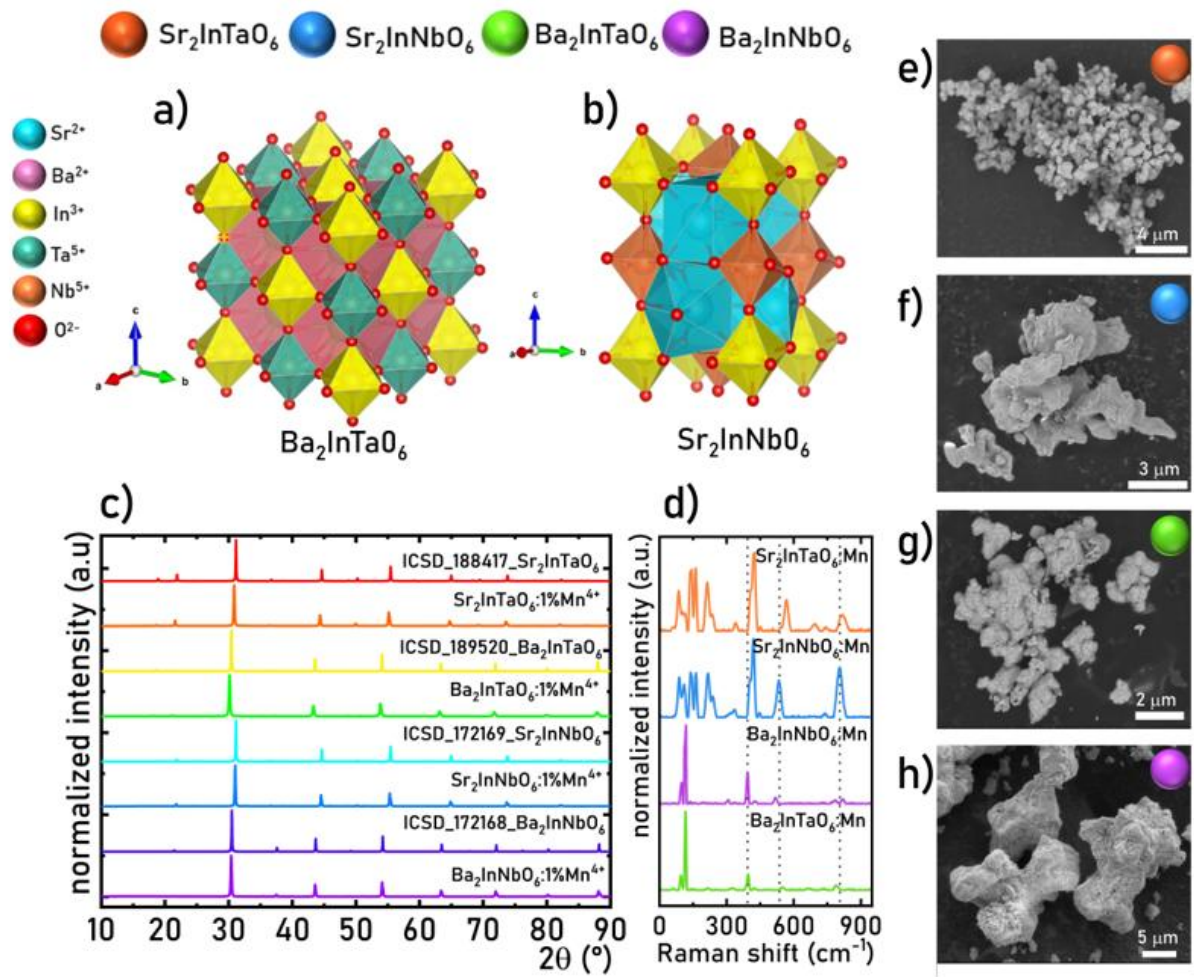

**Figure 1**. Graphical representation of the cubic structure of $Ba_2InTaO_6$ - a) and monoclinic structure of $Sr_2InNbO_6$ - b); comparison of room temperature X-ray diffraction patterns of $Sr_2InTaO_6$, $Ba_2InTaO_6$, $Sr_2InNbO_6$, and $Ba_2InNbO_6$ doped with 1%$Mn^{4+}$ ions with reference patterns - c); room temperature Raman spectra of the analyzed compounds – d); representative SEM images of $Sr_2InTaO_6:Mn^{4+}$– e), $Sr_2InNbO_6:Mn^{4+}$ - f), $Ba_2InTaO_6:Mn^{4+}$ - g), $Ba_2InNbO_6:Mn^{4+}$ - h).

The overall optical behaviour of the $Mn^{4+}$ center can be interpreted using the schematic configurational-coordinate diagram shown in Figure 2a. Because $Mn^{4+}$ ions experiences a strong octahedral crystal field arising from its high effective charge, its luminescence is characteristically dominated by sharp-line emission from the spin-forbidden $^2E_g \rightarrow ^4A_{2g}$ transition (Figure 2b, c). The low temperature (T= 83 K) excitation spectra of the $Mn^{4+}$-doped double perovskites presented in Figure 2d reveal the presence of four major bands: the host lattice absorption (HL), the $O^{2-} \rightarrow Mn^{4+}$ charge transfer transition (CT), and the spin-allowed but parity-forbidden $^4A_2 \rightarrow {}^4T_1$ and $^4A_2 \rightarrow {}^4T_2$ transitions (Table 2). Based on these values the crystal field parameters can be determined. The energy separation between the $^4A_2$ ground and $^4T_2$ excited states defines the octahedral crystal field strength, $\Delta_o = 10Dq$. The crystal field parameter $Dq$ is obtained by the formula: $E(^4A_2 \rightarrow {}^4T_2)/10 = Dq$, where $E(^4A_2 \rightarrow {}^4T_2)$ is the peak energy of the optical transition. As discussed earlier, the point charge model predicts $Dq$ to vary as $R^{-5}$, with $R$ representing the $Mn^{4+}$-$O^{2-}$ bond distance[77]. Analysis of the recorded spectra reveals that substitutions at both the $A$ and $B`$ sites affect the energy of the $Mn^{4+}$ levels. Replacement of $Ba^{2+}$ with $Sr^{2+}$ at the $A$ site induces a pronounced blueshift of the $^4A_2 \rightarrow {}^4T_1$ band, corresponding to energy differences of 1771 cm$^{-1}$ for the $Nb^{5+}$-based hosts and 1888 cm$^{-1}$ for the $Ta^{5+}$-based analogues. Consistent with their shorter ($Nb^{5+}$/$Ta^{5+}$)-$O^{2-}$ bond lengths (Table 1), the $Sr^{2+}$-based analogs exhibit larger $Dq$ values. In contrast, substitution at the $B'$ site ($Nb^{5+} \rightarrow Ta^{5+}$) results in considerably smaller spectral changes, manifested by a slight redshift of approximately 120 cm$^{-1}$ in the $Sr^{2+}$-based compounds, while having a negligible effect on the $Ba^{2+}$-based materials.

**Table 2**. Energies of the $Mn^{4+}$ transitions and $Dq$ determined based on the excitation spectra measured at 83K for the analyzed host materials.

| host | HL (cm$^{-1}$) | CT (cm$^{-1}$) | $^4A_2 \rightarrow {}^4T_1$ (cm$^{-1}$) | $^4A_2 \rightarrow {}^4T_2$ (cm$^{-1}$) | $Dq$ (cm$^{-1}$) | ($Nb^{5+}$/$Ta^{5+}$)-$O^{2-}$ (Å) |
|---|---|---|---|---|---|---|
| Sr$_2$InTaO$_6$ | | 30 148 | 26 346 | 18 018 | 1802 | 2.0106 |
| Sr$_2$InNbO$_6$ | | 29 736 | 26 227 | 17 967 | 1796 | 2.003 |
| Ba$_2$InTaO$_6$ | | 29 469 | 24 458 | 16 547 | 1655 | 2.016 |
| Ba$_2$InNbO$_6$ | | 28 943 | 24 456 | 16 728 | 1673 | 2.0705 |

The emission spectra recorded at 83 K consists of the zero-phonon line (*R*-line) along with the Stokes and anti-Stokes vibronic bands (Figure 2b, c). Comparison of emission spectra reveals that substituting the *A*- and *B'*-site cations induces modifications to the crystal-field environment surrounding the $Mn^{4+}$ ions, manifested by differences in optical response (Figure 2c). To further analyze the optical data, the values of the Racah parameters *B* and *C* were calculated from the experimental excitation and emission spectra using the following relationships (Table 3)[28,33]:

$$B = \frac{\left(\frac{\Delta E_T}{Dq}\right)^2 - 10\left(\frac{\Delta E_T}{Dq}\right)}{15\left(\frac{\Delta E_T}{D_q} - 8\right)} Dq \quad (5)$$

$$\Delta E_T = E(^4T_{1g}) - E(^4T_{2g}) \quad (6)$$

$$C = 0.328 \Delta E_E - 2.59 B + 0.59 \frac{B^2}{Dq} \quad (7)$$

$$\Delta E_E = E(^2E_g) \quad (8)$$

The nephelauxetic $\beta_1$ parameter was calculated using the following equation[78]:

$$\beta_1 = \sqrt{\left(\frac{B}{B_0}\right)^2 + \left(\frac{C}{C_0}\right)^2} \quad (9)$$

where $B_0$ and $C_0$ represent the Racah parameters for $Mn^{4+}$ ion in free state ($B_0$=1160 cm$^{-1}$, $C_0$ =4303 cm$^{-1}$)[78].

**Table 3**. The Racah parameters and spectroscopic parameters of $Mn^{4+}$ ions in the analyzed host materials.

| host | *B* (cm$^{-1}$) | *C* (cm$^{-1}$) | $^2$E (cm$^{-1}$) | $\beta_1$ | *Ri* =ZPL/$\nu_6$ | $\tau_{avr}$ (ms) |
|---|---|---|---|---|---|---|
| Sr$_2$InTaO$_6$ | 890 | 2806 | 14 777 | 1.007 | ~0.42 | 0.181 |
| Sr$_2$InNbO$_6$ | 875 | 2839 | 14 776 | 1.002 | ~0.43 | 0.164 |
| Ba$_2$InTaO$_6$ | 855 | 2762 | 14 364 | 0.977 | ~0.48 | 0.033 |
| Ba$_2$InNbO$_6$ | 820 | 2815 | 14 320 | 0.963 | ~0.71 | 0.014 |

As can be observed from the data assembled in Table 3, the energy of the $^2$E state is lowered by about 430 cm$^{-1}$ when moving from Sr$_2$In(Ta/Nb)O$_6$:Mn$^{4+}$ to Ba$_2$In(Ta/Nb)O$_6$:Mn$^{4+}$. The value of the $\beta_1$ parameter is also lower in the Ba$^{2+}$-phases relative to their Sr$^{2+}$-counterparts. As discussed earlier, the energy of the $^2$E state depends on the Mn$^{4+}$-O$^{2-}$ bonding covalence. Further, higher the covalence, lower is the value of the $\beta_1$ parameter. This shows that the covalent bonding is much stronger in the perovskite framework of the Ba$^{2+}$-phases than their Sr$^{2+}$-counterparts. The difference in covalency between the Ba$^{2+}$- and Sr$^{2+}$-based perovskite frameworks can be rationalized by considering the geometry of their octahedral units. In an ideal octahedron, all six Mn$^{4+}$-O$^{2-}$ bonds are equivalent, and the 15 characteristic angles consist of three linear (180°) and twelve right angles (90°). Distortions in a perovskite octahedron arise either from unequal Mn$^{4+}$-O$^{2-}$ bond lengths or from deviations of the octahedral angles from the ideal 90° value. In the Ba$^{2+}$ phases, the octahedra remain essentially ideal. This highly symmetric environment enhances Mn$^{4+}$ t$_{2g}$-O$^{2-}$ p$_\pi$ orbital overlap, leading to increased Mn$^{4+}$-O$^{2-}$ covalency even though the bond distances are longer. In contrast, the Sr$^{2+}$ phases exhibit significant octahedral distortion and tilting, with In$^{3+}$-O$^{2-}$-(Ta,Nb)$^{5+}$ angles reduced to ~169°, producing a distorted framework. The deviation from ideal symmetry diminishes the Mn$^{4+}$ t$_{2g}$-O$^{2-}$ p$_\pi$ orbital overlap and consequently lowers the covalency of the Mn$^{4+}$-O$^{2-}$ bond.

It is noteworthy that $\beta_1$ and the energies of the $^2$E-state in Sr$_2$InNbO$_6$:Mn$^{4+}$ and Sr$_2$InTaO$_6$:Mn$^{4+}$ are very similar. Given the Pauling electronegativities of Nb$^{5+}$ (1.6) and Ta$^{5+}$ (1.5), one would normally expect the (InNbO$_6$) framework to be slightly more covalent than (InTaO$_6$). This expectation is supported by the lower-energy O$^{2-}$→Mn$^{4+}$ charge-transfer band observed in Ba$_2$InNbO$_6$ (29 736 cm$^{-1}$) compared with its value of 30 148 cm$^{-1}$ in Sr$_2$InTaO$_6$.

Nevertheless, both the $^2$E-state energy and the covalence parameter $\beta_1$ remain nearly identical in these two $Sr^{2+}$-based double perovskites. In contrast, the $Ba^{2+}$ analogues display a clearer trend: the (InNbO$_6$) framework is indeed more covalent than its $Ta^{5+}$-containing counterpart, consistent with the electronegativity difference. This is further evidenced by the lower-energy of $O^{2-}\rightarrow Mn^{4+}$ charge-transfer band in $Ba_2InNbO_6$ (28 943 cm$^{-1}$) relative to $Ba_2InTaO_6$ (29 469 cm$^{-1}$). Among all compositions examined, $Ba_2InNbO_6$:$Mn^{4+}$ exhibits the highest $Mn^{4+}$-$O^{2-}$ bond covalency. The compositional dependence of the Racah parameter $B$ and the $Dq/B$ and $\beta_1$ parameter are visualized in Figure 2f, 2g and 2h, respectively. The analysis of the presented data reveals that the $Dq/B$ values follow the trend, $Ba_2InTaO_6$:$Mn^{4+}$ ($Dq/B$=1.93) < $Sr_2InTaO_6$:$Mn^{4+}$ ($Dq/B$=2.02) < $Ba_2InNbO_6$:$Mn^{4+}$ ($Dq/B$=2.04) < $Sr_2InNbO_6$:$Mn^{4+}$ ($Dq/B$=2.05). With the exception of $Ba_2InNbO_6$, $Dq/B$ decreases with increasing $Mn^{4+}$-$O^{2-}$ distances, consistent with the Tanabe-Sugano diagram. We note that the deviation observed for $Ba_2InNbO_6$, which has the longest $Mn^{4+}$-$O^{2-}$ bond distances among all hosts considered, may arise from its markedly higher covalency. This compound exhibits the smallest $B$ and $\beta_1$ value of any composition in this study.

The comparison of luminescence decay profile of $Mn^{4+}$ in $Sr_2InTaO_6$, $Sr_2InNbO_6$, $Ba_2InTaO_6$ and $Ba_2InNbO_6$, measured at 83 K shown in Figure 2e reveals that the $Ba^{2+}$-based compounds exhibit noticeably shorter $\tau_{avr}$ values than their $Sr^{2+}$-based analogs, indicating that $\tau_{avr}$ is strongly dependent on the host lattice. In general, several structural and electronic factors can influence the decay kinetics of $Mn^{4+}$ ions, and these aspects are discussed in detail later in this work. Table 4 summarizes the peak intensity ratio of the $R$-line to its most intense Stokes $v_6$ sideband ($Ri$ ). A striking aspect of these data is the unusually high $R$-line intensity observed for $Ba_2InNbO_6$ (~ 0.71), which is markedly greater than that of the $Ta^{5+}$-analog (~0.45). The substituted $B'$ site ($Ta^{5+}$/$Nb^{5+}$) possesses inversion symmetry not only with respect to the nearest-neighbour oxygen coordination but also relative to the next-nearest-neighbour cations.

Consequently, no static odd-parity distortions are present, and the *R*-line is a magnetic-dipole transition with low oscillator strength. In contrast, the vibronic sidebands, activated by odd-parity lattice vibrations, are electric-dipole allowed and therefore expected to dominate the spectrum. Under such centrosymmetric coordination, the ZPL/$v_6$ ratio should remain small. Yet, in $Ba_2InNbO_6$, the *R*-line exhibits a surprisingly large gain in intensity, deviating from this expectation.

As previously shown in these double perovskites, *B*-site cation mixing removes the inversion symmetry at the $Mn^{4+}$ center with respect to the next-nearest-neighbour cations. This symmetry breaking relaxes the parity selection rule and leads to an increase in the *Ri* value. In $Ba_2InTaO_6$, the degree of interchange among the *B*-site cations is estimated to be about 25% [56,67]. However, this mechanism cannot account for the difference in *Ri* observed between $Ba_2InNbO_6$ and $Ba_2InTaO_6$, because the extent of *B*-site mixing, which is driven primarily by the ionic-radius and charge differences between the *B*-site cations, is the same in both compositions. Therefore, this difference can be attributed to the greater covalency of $Ba_2InNbO_6$ relative to $Ba_2InTaO_6$. The enhanced covalence in the $Nb^{5+}$-analog arises from the stronger hybridization between the $Nb^{5+}$ $t_{2g}$ orbitals and the $O^{2-}$ $2p\pi$ orbitals. Although both $Nb^{5+}$ ($4d^0$) and $Ta^{5+}$ ($5d^0$) possess empty orbitals capable of forming covalent interactions with the oxide $2p$ orbitals, the $Nb^{5+}$-$O^{2-}$ hybridization is more effective. In addition, contributions from the $In^{3+}$ $5s$ orbitals also play a non-negligible role in strengthening the overall covalent interaction. The *Ri* value does not change between $Sr_2InNbO_6$ and $Sr_2InTaO_6$ due to the fact that in these compositions the bonding covalence within the perovskite framework is the same as evidenced by the similar energy of the $^2E$ state and the value of the $\beta_1$ parameter. In $Sr_2InTaO_6$, the degree of interchange among the *B*-site cations is estimated to be about 10%, which is less than that observed in the corresponding $Ba^{2+}$-analog[47,67]. This is expected based on the Woodward observation that the mixing between the *B*-site cations decreases with decreasing ionic radii of the *A* cation and lowering of the symmetry. The lower covalence of the perovskite framework in the $Sr^{2+}$-phases and the smaller degree of *B*-site mixing is responsible for the much smaller $Ri$ in the $Sr^{2+}$-phases relative to the $Ba^{2+}$-based counterparts.

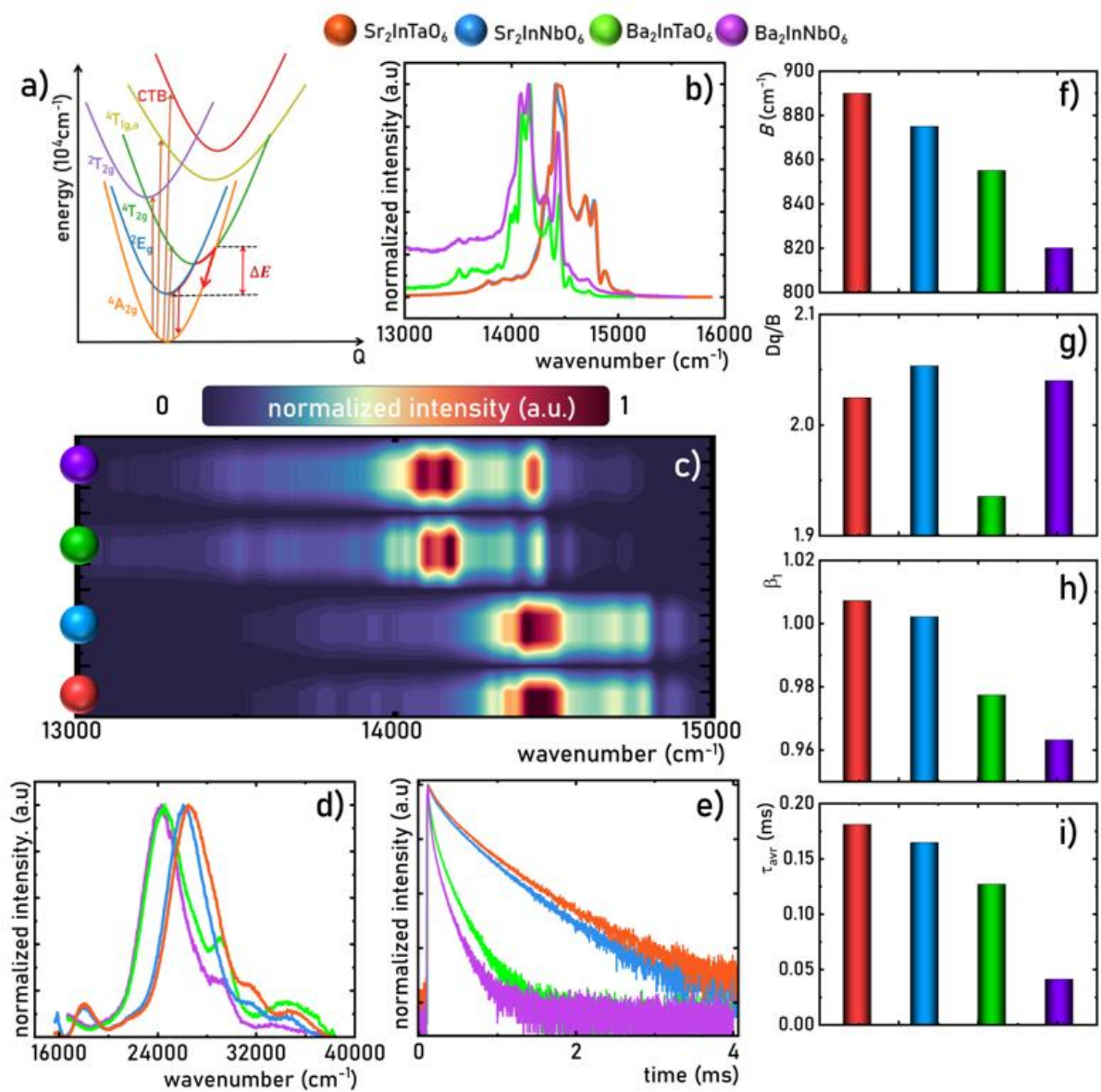

**Figure 2.** Simplified configurational-coordination diagram of $Mn^{4+}$ ions – a); the comparison of normalized emission spectra of analyzed phosphors measured at 83 K – b); and corresponding maps – c); comparison of excitation spectra of analyzed phosphors measured at 83 K – d); and luminescence decay profiles – e); the values of $B$ – f), $Dq/B$ – g), $\beta_1$ – h) and $\tau_{avr}$ – i) for the analyzed phosphors.

To evaluate how crystal structure and chemical composition of the phosphor influence the thermal stability of $Mn^{4+}$ luminescence in the investigated double perovskites, temperature-dependent emission spectra were recorded. A representative set of emission spectra as a function of temperature for $Ba_2InNbO_6:Mn^{4+}$ is shown in Figure 3a, while the comparison of the thermal luminescence maps for all analyzed phosphors is shown in Figure 3b-e (see also Figure S1-4). Based on the thermal dependence of the integrated $Mn^{4+}$ emission intensity in the investigated double perovskites (Figure 3f), the temperature at which the intensity decreases to 50% of the low temperature value ($T_{50}$) was estimated (Figure 3g). The main mechanism responsible for thermal depopulation of the emitting state of $Mn^{4+}$ ions is associated with the intersection point between the $^2E$ and $^4T_2$ states parabola. Therefore, to analyze the thermal stability of the $Mn^{4+}$ luminescence, the activation energy for thermal quenching ($E_a$) was calculated using the relation:

$$\ln\left(\frac{I_0}{I}\right) = \ln(A) - \frac{E_a}{KT} \qquad (11)$$

where $I_0$ and $I$ are the integrated intensities of the emission spectra at lowest temperature and given temperature T, respectively, $A$ is a constant and $k$ refers to the Boltzmann constant (Figure 3h). The obtained data shows that both $E_a$ and $T_{50}$ follow the trend $Ba_2InNbO_6 < Ba_2InTaO_6 < Sr_2InNbO_6 < Sr_2InTaO_6$. Thermal quenching is therefore more pronounced in the $Ba^{2+}$-based compounds compared with their $Sr^{2+}$-based counterparts. Among all investigated compositions, $Ba_2InNbO_6:Mn^{4+}$ exhibits the most rapid loss of emission intensity with increasing temperature, whereas $Sr_2InTaO_6:Mn^{4+}$ displays the highest thermal stability. Within each isostructural pair, the $Ta^{5+}$-containing analogues show higher $E_a$ and $T_{50}$ values than the corresponding $Nb^{5+}$-based compositions. As will be explained later in the text these results can be rationalized based on the energy separation between the zero-phonon levels of the $^2E$ and $^4T_2$ states since thermal quenching proceeds via the $^2E \rightarrow {}^4T_2 \rightarrow {}^4A_2$ crossover mechanism.

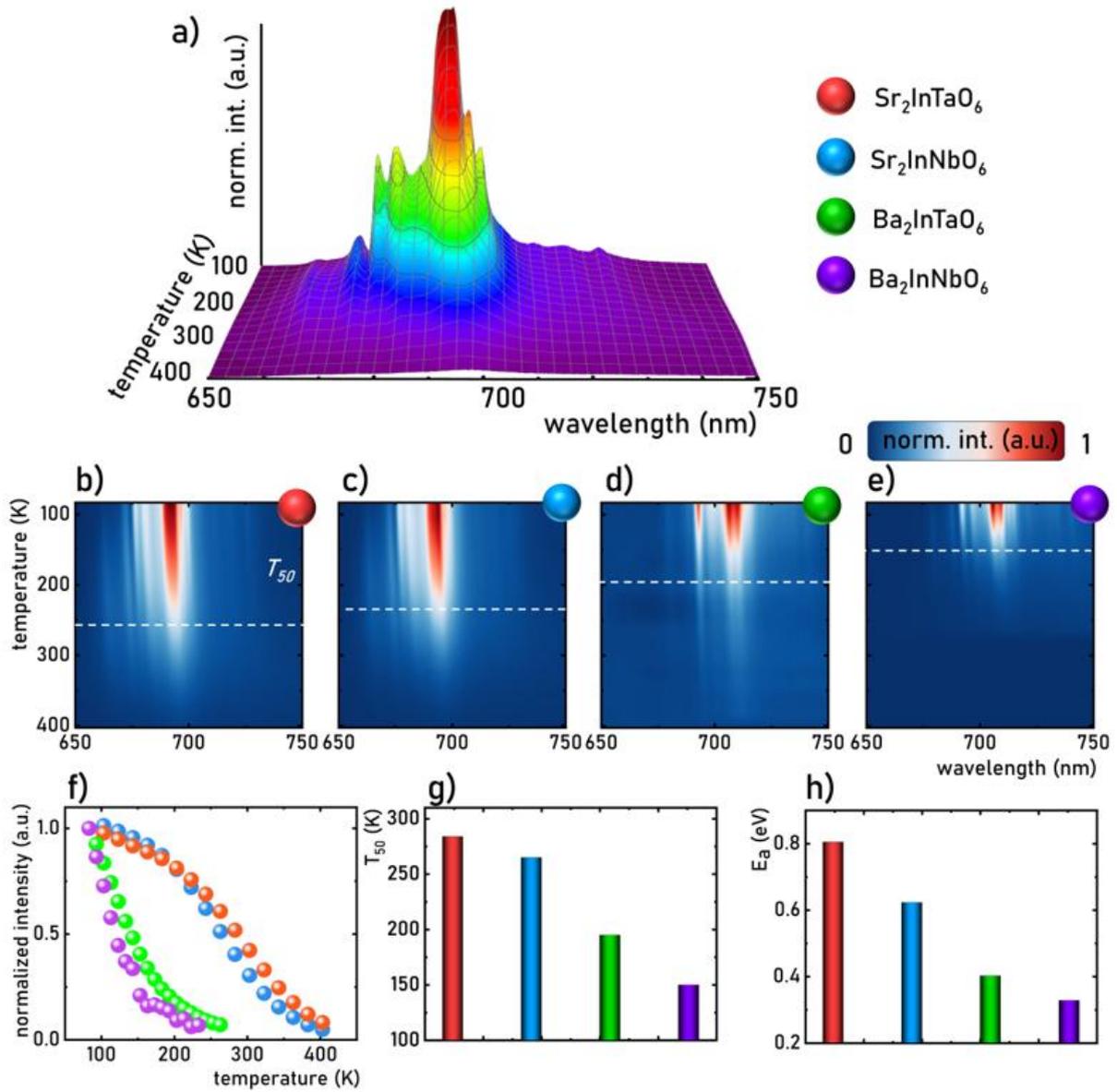

**Figure 3.** Thermal evolution of emission spectra of $Ba_2InNbO_6$:$Mn^{4+}$ - a); thermal maps of normalized emission spectra of $Sr_2InTaO_6$:$Mn^{4+}$ -b); $Sr_2InNbO_6$:$Mn^{4+}$ -c); $Ba_2InTaO_6$:$Mn^{4+}$ -d); $Ba_2InNbO_6$:$Mn^{4+}$ -e); thermal dependence of integral emission intensities of $Mn^{4+}$ ions in the analyzed host materials – f); corresponding $T_{50}$ – g); and $E_a$ -h).

The thermal quenching of the $Mn^{4+}$ emission intensity is expected to influence the luminescence kinetics of the $^2E$ state. At low temperatures, the decay is characterized by long $\tau_{avr}$ values, reflecting the slow nonradiative relaxation of electrons in the $^2E_g$ state (Figure 4a, Figure S5-7). As the temperature increases, $\tau_{avr}$ becomes shorter, indicating that thermal activation enhances nonradiative pathways. Consistent with the previously determined activation energies ($E_a$), the $Ba^{2+}$-based compounds exhibit very rapid thermal shortening of $\tau_{avr}$ (Figure 4b). In contrast, the $Sr^{2+}$-based materials show smoother decay curves, with $\tau_{avr}$ decreasing gradually with temperature, in agreement with their emission-intensity trends. For the $Ba^{2+}$ compounds, the $\tau_{avr}$ profile is sensitive to changes at the $B'$ site, whereas such $B'$-site variations have little effect on the thermal evolution of $\tau_{avr}$ in the $Sr^{2+}$ analogues. Notably, $Ba_2InNbO_6:Mn^{4+}$ shows almost negligible temperature-induced variation in $\tau_{avr}$. To quantify the thermal shortening of $\tau_{avr}$ in the analyzed phosphors the absolute $S_A$ and relative sensitivity $S_R$ were calculated as follows:

$$S_A = \frac{\Delta \tau}{\Delta T} \qquad (11)$$

$$S_R = \frac{1}{\tau}\frac{\Delta \tau}{\Delta T} \times 100\% \qquad (12)$$

where $\Delta\tau_{avr}$ represents the change in the $\tau_{avr}$ corresponding to the change of temperature by $\Delta T$. In the case of all analyzed phosphors a single dominant maximum in $S_A$ can be found, which value and the corresponding temperature at which it occurs are tunned with the modification in the host material composition (Figure 4c). The $S_{AMAX}$ determined for the investigated phosphors are as follows: 1.4 µs K$^{-1}$ at 283 K for $Sr_2InTaO_6:Mn^{4+}$, 1.165 µs K$^{-1}$ at 242 K for $Sr_2InNbO_6:Mn^{4+}$, 0.428 µs K$^{-1}$ at 222 K for $Ba_2InTaO_6:Mn^{4+}$ and 0.112 µs K$^{-1}$ at 108 K for $Ba_2InNbO_6:Mn^{4+}$ (Figure 4d). In the case of $S_R$, the obtained results revealed that $Ba_2InTaO_6:Mn^{4+}$ and $Ba_2InNbO_6:Mn^{4+}$ reach maximum $S_R$ values of 1.0 % K$^{-1}$ at 230 K and 0.49 % K$^{-1}$ at 110 K, respectively, after which their $S_R$ values decrease sharply (Figure 4e). In contrast, $Sr_2InNbO_6:Mn^{4+}$ exhibits an $S_R$ =1.2 % K$^{-1}$ at 280 K, while $Sr_2InTaO_6:Mn^{4+}$ reaches $S_R$ =1.8 % K$^{-1}$ at 307 K (Figure 4f). Moreover, when comparing the temperature at which the $S_R$ reaches its maximum ($T@S_{max}$) in relation to changes at the $B'$ site, a correlation with the average $Mn^{4+}$-$O^{2-}$ distance was observed. For $Sr^{2+}$-based compounds, increasing the crystal field strength by substituting Ta$^{5+}$ to Nb$^{5+}$ at the *B'* site leads to a decrease in *T@S$_{max}$*. A similar trend is observed in the Ba$^{2+}$-based hosts, where replacing Ta$^{5+}$ with Nb$^{5+}$ likewise leads to a reduction in *T@S$_{max}$*.

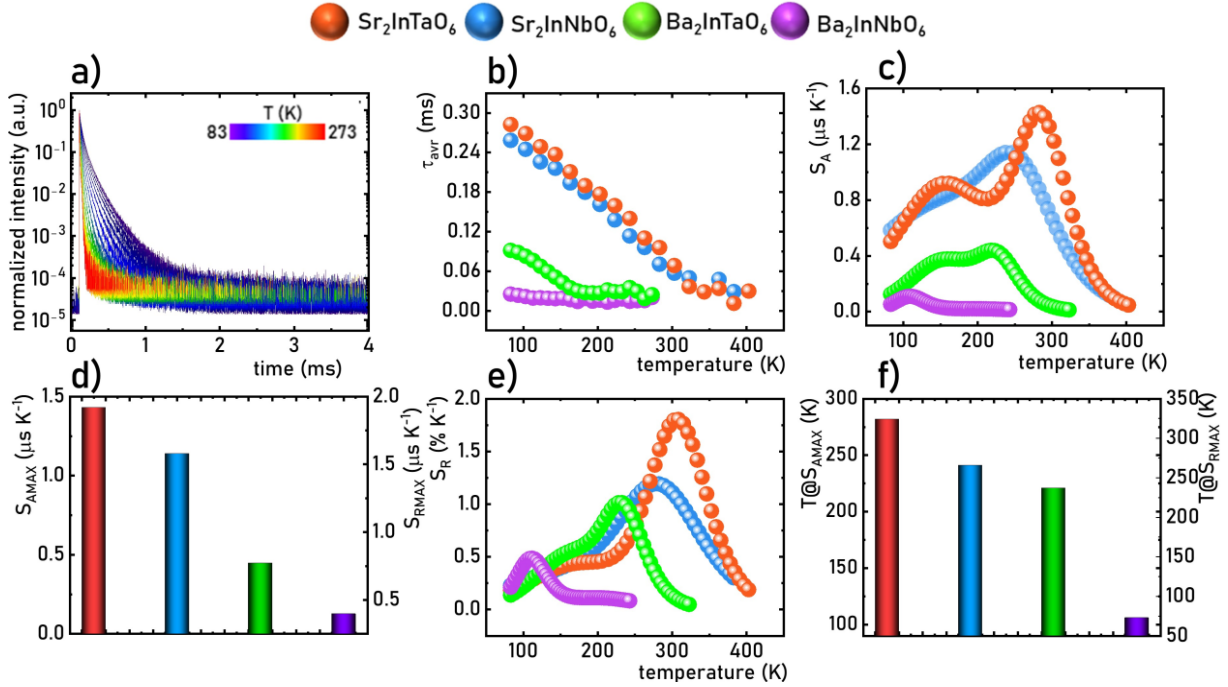

**Figure 4**. Luminescence decay profile of the Mn$^{4+}$ ions in Ba$_2$InTaO$_6$:Mn$^{4+}$ measured as a function of temperature– a); thermal dependence of $\tau_{avr}$ for analyzed phosphors - b) and corresponding $S_A$ - c), the $S_{AMAX}$ for different phosphors – d); thermal dependence of $S_R$ – e) and $S_{RMAX}$ for different host material compositions – f).

To establish a methodology for designing lifetime-based luminescent thermometers employing Mn$^{4+}$ ions, it is essential to identify the factors governing their relative sensitivity and to elucidate how these factors are modulated by the structural parameters of the host lattice. The most important thermometric parameter, which is the relative sensitivity $S_R$ referenced also in Equation 11, can alternatively be expressed in a form that utilize $S_A$, namely:

$$S_R = \frac{1}{\tau}\frac{\Delta \tau}{\Delta T} \times 100\% = \frac{1}{\tau} S_A \qquad (13)$$

From this expression, it is evident that two principal aspects must be addressed when thinking about the modification of the sensitivity of the lifetime-based thermometer: the dependence of the lifetime $\tau_{avr}$ on the host material, and the influence of the host material on the temperature-induced dynamics of $\tau_{avr}$. The general expression describing the probability of depopulation of the excited state includes contributions from both radiative and nonradiative processes, as follows[79]:

$$\frac{1}{\tau} = W_R + W_{NR} = \frac{1}{t_0} + W_{NR} \qquad (14)$$

At low temperatures, the probability of nonradiative depopulation ($W_{NR}$) is relatively small for $Mn^{4+}$ ions due to the high activation barrier associated with the nonradiative relaxation of the $^2E$ state. Consequently, the influence of the host material on $\tau_R$ becomes a major factor in these considerations. A preliminary inspection of the $S_R$ expression indicates that, from the perspective of maximizing sensitivity of luminescence thermometer, it is advantageous for the $^2E$ state to exhibit a short $\tau_{avr}$. In transition metal ions with a $3d^3$ electronic configuration, such as $Mn^{4+}$, the spin-orbit coupling between the $^2E$ and $^4T_2$ states exerts a strong and dominant effect on the radiative lifetime $\tau_R$. Therefore, $\tau_R$ strongly depends on the energy gap between the $^2E$ and $^4T_2$ levels ($\Delta$) and the lifetime of the $^4T_2$ state ($\tau_T$)[80]. Therefore, the radiative lifetime of the $^2E$ doublet state depends primarily on the doublet-quartet energy separation ($\Delta$). Figure 5a presents the values of $\Delta$ for the investigated double perovskites. The $Sr^{2+}$-based compounds exhibit larger $\Delta$ values than their $Ba^{2+}$-based analogues. This may be related with the shorter average $Mn^{4+}$-$O^{2-}$ bond distances and correspondingly higher crystal-field strength ($Dq$) in the $Sr^{2+}$ phases. Substituting $Ta^{5+}$ with $Nb^{5+}$ produces only a modest decrease of ~50 cm$^{-1}$ in for the $Sr^{2+}$ compounds, whereas in the $Ba^{2+}$ compounds the same substitution increases $\Delta$ by ~225 cm$^{-1}$. These contrasting trends indicate that the octahedral crystal-field strength and $Mn^{4+}$-$O^{2-}$ bonding covalence are similar within the $Sr^{2+}$ series but differ substantially within the $Ba^{2+}$ series. This is not surprising since $Ba_2InNbO_6$ exhibits the highest $Mn^{4+}$-$O^{2-}$ covalence among the compositions studied. In this case, enhanced covalence lowers the energy of the $^2E$ level and raises that of the $^4T_2$ level, thereby increasing $\Delta$ (Table 2). This explains why $\Delta(Ba_2InTaO_6; 2183\ cm^{-1})$ is smaller than $\Delta(Ba_2InNbO_6; 2408\ cm^{-1})$. The dependence of the $\tau_R$ on the energy separation $\Delta$ shown in Figure 5b reveals that the $Sr^{2+}$-based compounds exhibit longer $\tau_R$ values than the $Ba^{2+}$-based analogues, consistent with the expectation from Eq. 14, since $\Delta$ is larger in the $Sr^{2+}$ compounds. However, despite the higher $\Delta$ value of $Ba_2InNbO_6$, its $\tau_R$ is significantly shorter than that of $Ba_2InTaO_6$. We attribute this behaviour to the stronger hybridization between the O $2p$ orbitals and the $Mn^{4+}$ $d$ orbitals in $Ba_2InNbO_6$, which relaxes the spin-forbidden nature of the $^2E_g \rightarrow {}^4A_{2g}$ transition. The increased transition probability also accounts for the higher intensity of the zero-phonon $R$-line observed in $Ba_2InNbO_6$. Figure 5c shows the variation of $\tau_R$ as a function of $Dq$. The $Sr^{2+}$-based compounds with shorter $Mn^{4+}$-$O^{2-}$ bond distance and therefore higher $Dq$ exhibits longer $\tau_R$ values than the $Ba^{2+}$-based analogues. Therefore, in general, the strength of the crystalline field can be used to tune the $\tau_R$ values. On the other hand, direct correlation between the energy of the $^2E$ state and the $\tau_R$ was not found for the analyzed host materials (Figure 5d). Since the energy of the $^4T_2$ state is also dependent on the covalency of the $Mn^{4+}$-$O^{2-}$ bond, the linear correlation was found between the $\beta_1$ parameter as a function of $\Delta$ (Figure 5e). Hence, as shown in Figure 5f, the higher is the bonding covalence, the shorter is $\tau_R$. This is again related to the relaxation of the spin selection rule by the enhanced hybridization between the $Mn^{4+}$ $d$ orbitals and the O $2p$ orbitals. Therefore, it becomes evident that, from the standpoint of achieving a short $\tau_R$, which may positively affect the relative sensitivity $S_R$ it is advantageous to employ host materials characterized by low $\beta_1$ values; that is, materials in which the $Mn^{4+}$-$O^{2-}$ bond exhibits a higher degree of covalency.

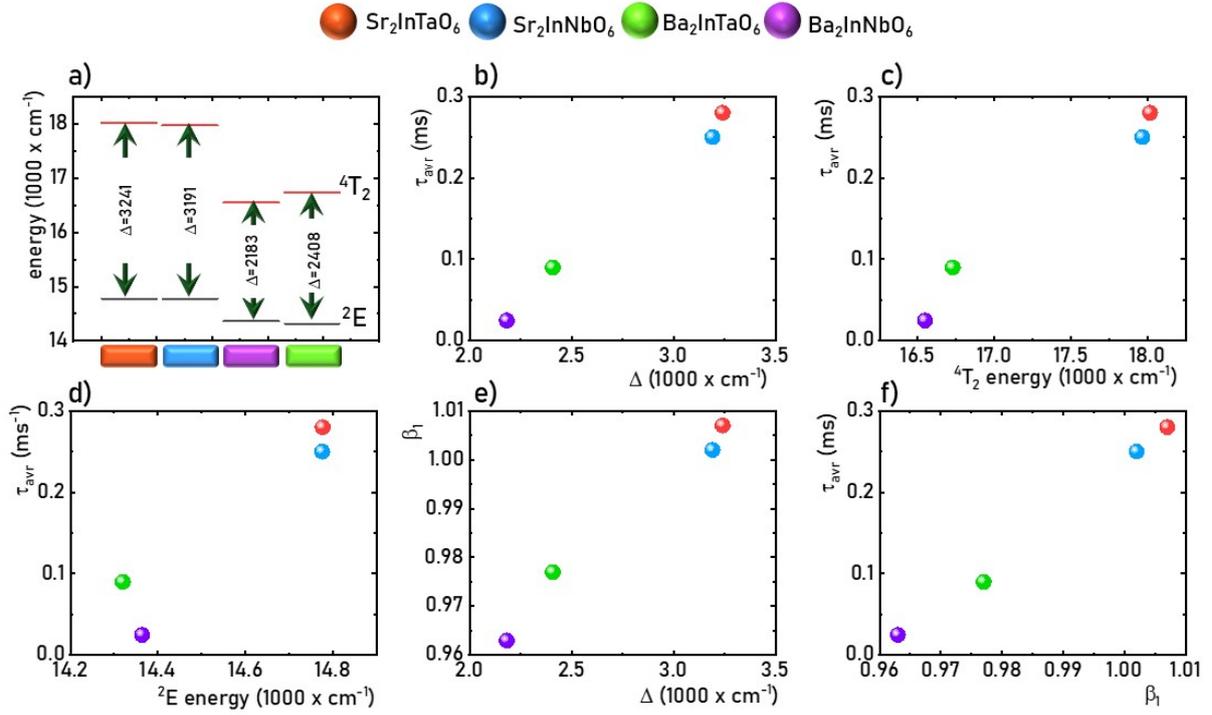

**Figure 5**. The Δ for the analyzed host materials – a); the influence of the Δ on the $\tau_{avr}$ – b); the influence of the energy of the $^4T_2$ state – c) and of the $^2E$ state – d) on the $\tau_{avr}$; the $\beta_1$ as a function of Δ – e) and the influence of the $\beta_1$ parameter on the $\tau_{avr}$ – f).

Another aspect influencing the $S_R$ is the thermal dependence of the $\tau_{avr}$. The general expression describing the processes governing the temperature-induced evolution of the $\tau_{avr}$ is given by[80]:

$$\tau_E = \frac{\tau_{stat}\left(1+\exp\left(\frac{-h\omega}{kT}\right)+3\exp\left(\frac{-\Delta}{kT}\right)\right)}{\left(1+\frac{\tau_{stat}}{\tau_{dyn}}\exp\left(\frac{-h\omega}{kT}\right)\right)\left(\left(\frac{V_S-O}{\Delta^t}\right)^2+3\exp\left(\frac{-\Delta}{kT}\right)\right)} \quad (15)$$

where, $\tau_{stat}$, $\tau_{dyn}$, $h\omega$, $\Delta$, $\Delta_t$, $V_{S-O}$ and $k$ represent radiative transition kinetic parameter generated by static odd parity crystal field and probability of radiative transition generated by odd parity crystal vibration, phonon mode of the host materials, real energy difference between $^2E$ and $^4T_2$ states, and the energy difference determined from excitation spectrum, spin-orbital coupling constant and Boltzmann constant, respectively.

Although this equation provides a comprehensive description of the underlying processes, its practical application is challenging due to its mathematical complexity. From a thermometric design perspective, it is particularly important to understand how material-specific parameters influence both $S_{AMAX}$ and the corresponding temperature $T@S_{AMAX}$, as these determine the suitability of a luminescent thermometer for a specific application. In the analysed phosphors, the value of $S_{AMAX}$ increases substantially with relatively small variations in the $\beta_1$ parameter (Figure 6b). This behaviour may be attributed to the strong dependence of the $^4T_2$ energy level and consequently the activation energy $E_a$ on the parameter $\beta_1$. In principle, $\Delta$ is not expected to exhibit thermal sensitivity; however, changes in the $^4T_2$ energy simultaneously modify $E_a$, and thermalization from the $^2E$ to the $^4T_2$ level is widely recognized as the dominant process governing the thermal sensitivity of $Mn^{4+}$ luminescence kinetics. In $Sr^{2+}$-based compounds, as the $B'$ site changes from $Ta^{5+}$ to $Nb^{5+}$, the $Mn^{4+}$-$O^{2-}$ bond length decreases, leading to a decrease in $B$ value due to stronger delocalization of $3d$ electrons, accompanied by a decrease in $\beta_1$. This indicates an increase in covalency, explaining why $Sr_2InNbO_6$:$Mn^{4+}$ exhibits greater covalent character than $Sr_2InTaO_6$:$Mn^{4+}$. In contrast, for $Ba^{2+}$ compounds, the decrease in $Mn^{4+}$-$O^{2-}$ bond length from $Nb^{5+}$ to $Ta^{5+}$ results in an increase in $\beta_1$, indicating greater ionicity in $Ba_2InTaO_6$ compared to $Ba_2InNbO_6$:$Mn^{4+}$. Similarly to $S_{AMAX}$, the $T@S_{AMAX}$ also increases markedly with small changes in $\beta_1$ (Figure 6b and 6c, respectively). The influence of the $^4T_2$ energy level on the thermal depopulation of the $^2E$ state and thus on its thermal stability can be further illustrated by examining the relationship between the $T_{50}$ parameter, which directly characterizes the thermal stability of $Mn^{4+}$ emission intensity, and $S_{AMAX}$ (Figure 6e). A key factor contributing to thermally stable luminescence is the structural rigidity imparted by ordered host structures. In the group of phosphors analyzed, the $Sr^{2+}$-based compounds, which have shorter $Mn^{4+}$-$O^{2-}$ bond length than $Ba^{2+}$-based hosts, exhibit higher thermal stability. This can be attributed to the smaller atomic size of $Sr^{2+}$ which causes contraction and slight geometric distortion at the $A$ site, thereby promoting greater cationic ordering at the $B'$ site. As a result, due to the smaller $A$ site ions and lower crystal symmetry, the degree of $B$-site cationic order ($In^{5+}$-$Nb^{5+}$ or $In^{5+}$-$Ta^{5+}$) in $Sr_2In(Nb/Ta)O_6$:$Mn^{4+}$ is expected to be higher than that in the cubic $Ba_2In(Nb/Ta)O_6$:$Mn^{4+}$ system, which accounts for the enhanced thermal stability of the $Sr^{2+}$ based hosts. The highly monotonic correlation shown in Figure 6e provides direct evidence for this relationship, although the detailed mechanism remains insufficiently understood. As expected, $T@S_{AMAX}$ is identical to $T_{50}$, since both parameters describe the same underlying thermal depopulation process of the $^2E$ state, a conclusion further supported by the linear dependence presented in Figure 6f.

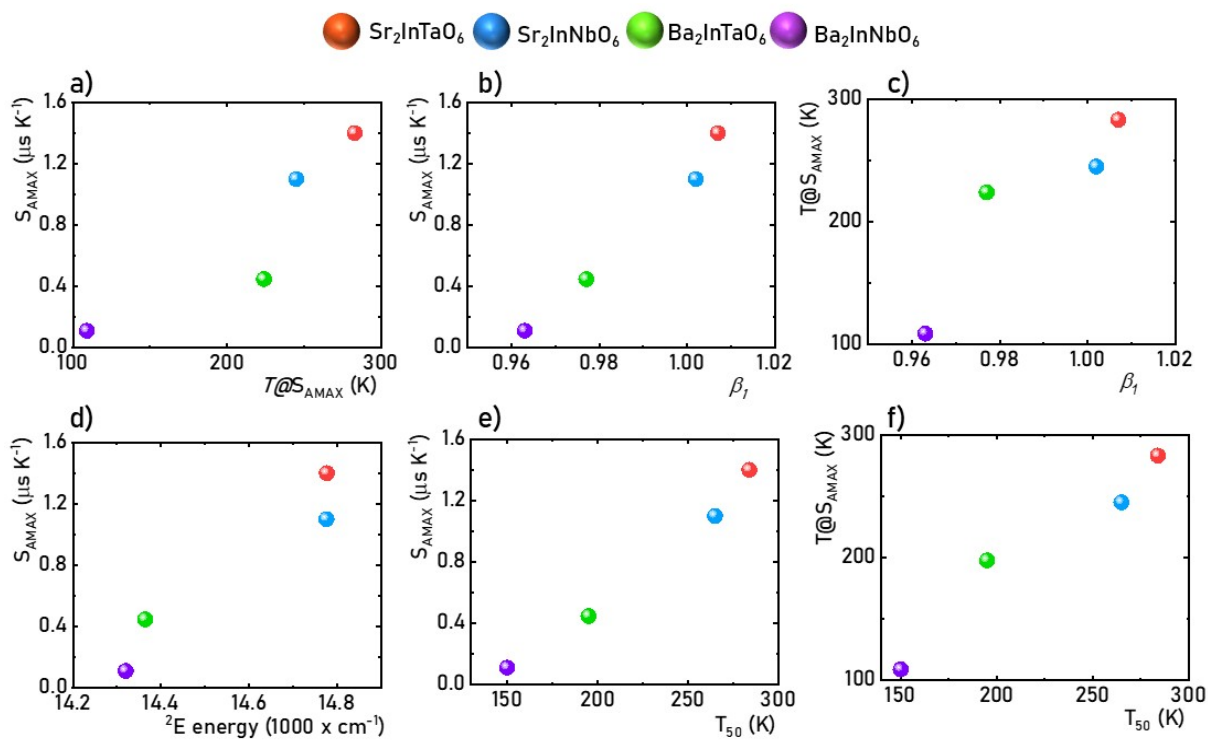

**Figure 6**. The influence of the $T@S_{AMAX}$ on the $S_{AMAX}$ – a); the $S_{AMAX}$ as a function of $\beta_1$ parameter – b) the influence of the $\beta_1$ parameter on the $T@S_{AMAX}$ – c); the dependence of the $S_{AMAX}$ on the energy of the $^2E$ state – d) and the $T_{50}$ - e); The $T@S_{AMAX}$ as a function of $T_{50}$ – f).

Since $S_R$ depends on both the $\tau_{avr}$ and $S_A$, establishing correlations between these parameters and the physicochemical characteristics of the host material is highly desirable. Notably, the $\beta_1$ parameter exerts a pronounced influence on both $\tau_{avr}$ and $S_{AMAX}$. This observation is particularly significant, as it was initially anticipated that the $Dq/B$ ratio would play the dominant role in governing the thermometric response. However, as demonstrated above, distortion of the crystallographic site symmetry occupied by $Mn^{4+}$ ions can substantially modify their spectroscopic properties. The clear linear correlations between $\tau_{avr}$ (Figure 7a) and $\beta_1$, as well as between $S_{AMAX}$ (Figure 7b) and $\beta_1$, combined with the absence of a direct relationship between these parameters and $Dq/B$ (Figure S8), indicate that $\beta_1$ is a more reliable descriptor for predicting the thermometric performance of lifetime-based luminescence thermometers. The dependence of $\tau_{avr}$ on $\beta_1$ is described by the experimental relationship as follows:

$$\tau_{avr}(\beta_1) = 5.91 \cdot \beta_1 - 5.6744 \tag{16}$$

On the other hand, $S_{AMAX}$ can likewise be expressed as a function of $\beta_1$:

$$S_{A\max}(\beta_1) = 28.112 \cdot \beta_1 + 26.988 \tag{17}$$

Consequently, according to Eq. 13, $S_{RMAX}$ may also be formulated as a function of $\beta_1$:

$$S_{R\max} = \frac{28.112 \cdot \beta_1 + 26.988}{5.91 \cdot \beta_1 - 5.6744} \tag{18}$$

As illustrated in Figure 7c, this experimentally derived relationship accurately reproduces the observed dependence of $S_{RMAX}$ on $\beta_1$. Although the equation is specific to the analyzed group of materials and requires further validation across a broader range of phosphors, it provides valuable guidance for the rational design of lifetime-based thermometers with targeted thermometric performance.

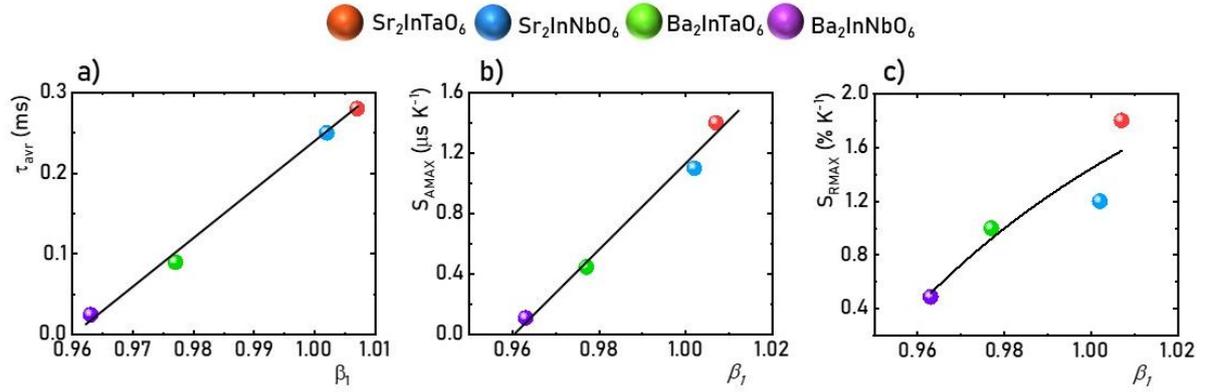

**Figure 7**. The influence of the $β_1$ on the $τ_{avr}$ – a), $S_{AMAX}$ – b) and $S_{RMAX}$ – c) for the analyzed group of $Mn^{4+}$ doped phosphors.

**Conclusions**

In summary, this work presents a systematic investigation of the influence of host composition on the thermometric performance of $Mn^{4+}$-activated lifetime-based luminescence thermometers in four double perovskite materials, namely $Sr_2InTaO_6:Mn^{4+}$, $Sr_2InNbO_6:Mn^{4+}$, $Ba_2InTaO_6:Mn^{4+}$ and $Ba_2InNbO_6:Mn^{4+}$. The $Sr^{2+}$-based phosphors crystallize in a monoclinic structure, whereas their $Ba^{2+}$-based counterparts adopt a cubic phase. In the $Ba^{2+}$-based hosts, $Mn^{4+}$ ions occupy nearly perfect octahedral sites. Substitution of $Ba^{2+}$ with $Sr^{2+}$ induces octahedral tilting and lowers the point symmetry of the $Mn^{4+}$-occupied sites. Structural analysis indicates that differences in ionic radii between $In^{3+}$ and $Nb^{5+}/Ta^{5+}$ promote $B/B'$ site disorder, which is more pronounced in the higher-symmetry $Ba^{2+}$-based compounds. Replacement of $Ba^{2+}$ by $Sr^{2+}$ at the $A$ site results in a blueshift of the $Mn^{4+}$ absorption bands, whereas substitution of $Nb^{5+}$ with $Ta^{5+}$ at the $B'$ site only marginally affects their spectral positions. The nephelauxetic parameter $β_1$ decreases for $Ba^{2+}$-based phosphors, indicating increased $Mn^{4+}$-$O^{2-}$ covalency, attributed to enhanced $Mn^{4+}$ $t_{2g}$–$O^{2-}$ pπ orbital overlap in the highly symmetric environment despite longer bond distances. Optimization analysis reveals that maximizing relative sensitivity ($S_R$) favours hosts with small $^2E$ - $^4T_2$ energy separation and low $β_1$ values, whereas the temperature sensitivity of the average lifetime ($S_{AMAX}$) increases with increasing $β_1$ and shifts to higher temperatures. Overall, thermometric performance is predominantly governed by the thermal evolution of $S_A$, suggesting that tuning this parameter is key to improving $Mn^{4+}$-based lifetime thermometers in double perovskites.

Although the conclusions drawn in this study are based on a limited set of host materials and require further systematic investigation to confirm the general validity of the proposed hypotheses, the design principles outlined herein provide valuable guidance for materials engineering. In particular, they offer strategic directions for the rational development of lifetime-based luminescence thermometers with predefined and optimized thermometric performance.

**Acknowledgements:**

This work was supported by the National Science Center (NCN) Poland under project no. DEC-2023/49/B/ST5/03384. Maja Szymczak gratefully acknowledges the support of the Foundation for Polish Science through the START program. Authors would like to acknowledge dr Damian Szymanski for SEM and EDS analyses.